\documentstyle[12pt,aaspp4,psfig]{article}
\tighten
\newcommand{\lsun}{\mbox{$L_\odot$}}

\begin{document}

\title {A Single Circumbinary Disk in the HD~98800 Quadruple System}

\author {D.W. Koerner\altaffilmark{1}, E.L.N.Jensen\altaffilmark{2}, 
K.L. Cruz\altaffilmark{1}, T.B. Guild\altaffilmark{1}, and 
K. Gultekin\altaffilmark{1}}
\altaffiltext{1} 
{University of Pennsylvania, David Rittenhouse Laboratory, 
209 S. 33rd St., Philadelphia, PA 19104-6396}
\altaffiltext{2}
{Dept. of Physics \& Astronomy, Swarthmore College, Swarthmore, PA 19081}

\bigskip
\bigskip

\begin{abstract}

We present sub-arcsecond thermal infrared imaging 
of HD~98800, a young quadruple system composed of 
a pair of low-mass spectroscopic binaries separated by 0.8$''$ 
(38 AU), each with a K-dwarf primary. Images at wavelengths
ranging from 5 to 24.5 $\mu$m show unequivocally that
the optically fainter binary, HD~98800B, is the sole source of a
comparatively large infrared excess upon which a silicate
emission feature is superposed. The excess is detected
only at wavelengths of 7.9~$\mu$m and longer, peaks at 25 $\mu$m,
and has a best-fit black-body temperature of 150~K, indicating
that most of the dust lies at distances  greater than the orbital
separation of the spectroscopic binary. We estimate 
the radial extent of the dust with a disk model 
that approximates radiation from the spectroscopic binary as a 
single source of equivalent luminosity.  Given the data, 
the most-likely values of disk properties in  
the ranges considered are
$R_{in} =  {5.0}\pm2.5$ AU, $\Delta R = 13\pm8$ AU,  
$\lambda_0 = {2}^{+4}_{-1.5}\mu$m, $\gamma = 0\pm2.5$,
and $\sigma_{total} = 16\pm3$ AU$^2$,  
where $R_{in}$ is the inner radius,
$\Delta R$ is the radial extent of the disk,  $\lambda_0$
is the effective grain size, $\gamma$ is the radial
power-law exponent of the optical depth, $\tau$, and $\sigma_{total}$
is the total cross-section of the grains. The range of implied 
disk masses is 0.001--0.1 times that of the moon.
These results show that, 
for a wide range of possible disk properties, a
circumbinary disk is far more likely than a narrow ring.

\end{abstract}

Subject headings: binaries: close --- binaries: spectroscopic --- 
                  circumstellar matter --- planetary systems --- 
                  stars: imaging --- stars: individual (HD 98800)

\vfill
\eject

\section {Introduction}

The evolution of circumstellar dust around young stars is 
traced by a time-dependent signature in excess infrared emission.
The evidence lies primarily in the spectral distribution of 
radiation as it chronicles infall from a protostellar 
envelope, viscous accretion in a gas-rich circumstellar disk, and 
dispersal in a dusty ``debris disk'' that survives as a 
last vestige of planet formation  
(Adams, Lada, \& Shu 1987; Backman \& Paresce 1993). 
Imaging has dramatically confirmed this interpretation,
providing support for a standard model of circumstellar evolution and 
elucidating the role of circumstellar disks throughout the process
(Beckwith \& Sargent 1996; Holland et al.\ 1998;
Koerner 1997; Koerner et al.\ 1998).  The
co-existence of disks with stellar companions is attested by
comparison of high-resolution binary surveys (Ghez, Neugebauer,  
\& Matthews 1993; Leinert et al.\ 1993)
with the results of imaging and long-wavelength flux measurements 
(Jensen et al.\ 1996a,b; Mathieu et al.\ 2000). 
Disks are found to be reduced in mass for
binaries with separations in the 10--100 AU range, similar to the
typical disk size. However,  
circumstellar disks in binaries wider than 100 AU
(Beckwith et al.\ 1990; Osterloh \& Beckwith 1995; Jensen et al.\ 1996a), 
and circum{\em binary} disks around spectroscopic binaries
(Jensen \& Mathieu 1997) are not obviously different  
from disks around single stars with respect to either their 
global properties or frequency of occurrence. 
These results argue strongly for the possibility of 
an abundant and diverse population of extra-solar planets.

Among pre--main-sequence spectroscopic binaries with 
separations of 1 AU or less, there is growing evidence that
circumbinary disks are common. Massive circumbinary disks have
been found in a handful of cases, demonstrating
unequivocally that the presence of a small-separation binary is not an
impediment to the formation of a protoplanetary disk (Jensen \&
Mathieu 1997).  Examples include
GW Ori (Mathieu et al.\ 1991, 1995), UZ~Tau~E (Jensen et al.\ 1996b;
Mathieu et al.\ 1996), and DQ Tau (Mathieu et al.\ 1997), all with
projected orbital separations of order 1 AU or
less and disk masses that are comparable to or greater than that
estimated for the minimum mass solar nebula.  

The degree of complexity possible for multiple star-disk
systems is perhaps nowhere better illustrated than in the 
case of the post-T Tauri quadruple system, HD~98800. 
It is composed of a pair of low-mass spectroscopic binaries, each with a
K-dwarf primary, that have a projected separation of 
0.8$''$ (37.6 AU at the 47 pc
distance determined by {\it Hipparcos}) and estimated ages
of $\sim10$~Myr (Soderblom et al.\ 1998). Despite the presence of many
stellar components, HD~98800 is associated with an unusually
strong IRAS signature of dust emission with a temperature similar
to the solar zodiacal dust bands (Walker \& Wolstencroft 1988;
Zuckerman \& Becklin 1993; Sylvester et al.\ 1996) and with
evidence for silicate emission from dust grains (Skinner, Barlow,
\& Justtanont 1992). Until recently, there were no observations
that provided a hint as to how this dust was distributed among the stellar
components of the system. N-band imaging that marginally
resolved the binary has now shown that most of the dust is
associated with the optical secondary and spectroscopic binary
HD~98800B (Gehrz et al.\ 1999). Here we present sub-arcsecond
images from 5 to 25 $\mu$m that fully resolve
the 0.8$''$ binary components of the HD~98800 system.

\section{Observations and Results}

HD~98800 was observed with JPL's mid-infrared camera MIRLIN at the f/40
bent-Cassegrain focus of the Keck II telescope on UT 14 March 1998.
MIRLIN employs a Boeing 128$\times$128 pixel, high-flux Si:As BIB detector
with a plate scale at Keck II of 0.137$''$ per pixel 
and 17.5$''$ field of view. Background subtraction
was carried out by chopping the secondary mirror at a $\sim$4 Hz rate with
8$''$ throw in the north-south direction, and by nodding the telescope a
similar distance east-west after coadding a few hundred chop
pairs. Images of the source on the 
double-differenced frames were shifted and added  
to make the final 32 $\times$ 32 
(4.4$''$ $\times$ 4.4$''$) images. Observations were
carried out at wavelengths from 4.7 to 24.5 $\mu$m in the
spectral bands listed in Table I.
Small dither steps were taken between chop-nod cycles. 
Infrared standards $\beta$ Leo (A3 V) and $\alpha$ Hya (K3 III)
were observed in the same way at similar airmasses.

The resulting images of HD~98800 are displayed in Fig.\ 1.  
Since the half-maximum width of the 
point spread function (PSF) is between 0.3$''$ and 0.55$''$ over the full 
wavelength range, it is possible to identify unambiguously the relative
flux densities of the mid-infrared emission for the first time.
Two point sources, separated by 0.81$''\pm0.02''$, are detected at 
wavelengths up to $\lambda$ = 12.5 $\mu$m with an orientation that
corresponds to the binary optical components with A to the south and
B to the north (cf.\ Soderblom et al.\ 1998). Only a single point source
is detected at the longest wavelengths.
It is immediately apparent from the images that this 
emission arises predominantly from the northern source, 
corresponding to the optical secondary HD~98800B. In contrast, emission
from the optical primary decreases steadily towards longer wavelengths.
Separate-component flux densities were derived by fitting a 
measured PSF to each component and using the resulting
flux component ratio to decompose the total flux into values 
for HD~98800A and B. 
Results are listed in Table I and plotted as a spectral energy distribution 
in Fig.\ 2 together with measurements from HST, IRAS, and the JCMT 
(Sylvester et al.\ 1996; Soderblom et al.\ 1998). Mid-infrared flux 
densities measured for the total system are in excellent agreement with 
earlier values published in the literature (Zuckerman \& Becklin 1993; 
Sylvester et al.\ 1996).

The distribution of mid-infrared flux between the components
HD~98800A and B clearly indicates that the total infrared 
excess of the system is dominated by the contribution from HD~98800B. 
Values for the flux density of HD~98800B at $\lambda$ = 12.5 and 24.5 $\mu$m 
agree very well with 12 and 25 $\mu$m IRAS fluxes measured for the whole 
system. In contrast, flux densities for HD~98800A
decrease approximately as $\lambda^{-2}$ between $\lambda$ = 7.9 and 
12.5 $\mu$m, consistent with origin in a stellar photosphere. At 
12.5 $\mu$m, emission from HD~98800A contributes less 
than 4\% of the total emission. At 24.5~$\mu$m, an
upper limit to its contribution comprises only
2\% and an estimated photospheric contribution only 0.2\% of the
total flux.
It is thus a good approximation to ascribe all the 
emission from unresolved measurements at $\lambda > 25\ \mu$m to HD~98800B 
and neglect any contribution from HD~98800A. This result is largely
in agreement with the conclusion of Gehrz et al.\ (1999), who
nevertheless attributed some of the mid-infrared excess emission
to HD~98800A on the basis of lower resolution imaging (1$''$ 
at $\lambda$ = 9.8 $\mu$m) which only marginally resolved the
0.8$''$ separation of components A and B.

A spectral signature of silicate emission 
at $\lambda$ $\approx$ 10~$\mu$m is evident in the flux 
measurements of HD~98800B plotted in Fig.\ 2. It is 
displayed in more detail in Fig.\ 3, where the measurements at 
7.9 and 12.5 $\mu$m have been assumed to represent featureless thermal 
continuum emission, and a simple linear extrapolation between the two
points has been subtracted off. The spectrum was then scaled to give 
the 7.9 and 12.5 um points a value of one to facilitate
comparison with other data from the literature (see Hanner, Lynch, \& 
Russell 1994 for comparison of different continuum removal techniques).  
Silicate features from comets and 
interstellar dust are plotted in Fig.\ 3 for comparison. 
It is readily apparent that the circumstellar dust feature 
resembles that from comets more than that from
the interstellar medium. The feature is broader and does not show 
a single narrow peak between 9 and 10 $\mu$m as seen for interstellar grains 
in the Trapezium. For comets, this broadened line-shape has been 
interpreted as diagnostic  of a mixture of amorphous and crystalline 
silicates that radiate predominantly at 9.8 and 11.2~$\mu$m, respectively
(Hanner, Lynch, \& Russell 1994).

\section {Modeling and Discussion} 


To better interpret the emission from HD~98800, we fit model emission
from stellar photospheres to 
optical (WFPC2) and near infrared (NICMOS) HST imaging that resolved
components A and B (Soderblom et al.\ 1998; Low et al.\ 1999). 
These were matched by reddened model atmospheres from Kurucz 
by varying only the stellar luminosity, as described by
Jensen \& Mathieu (1997). Discrepant 
NICMOS measurements at roughly the 
same wavelength were averaged and weighted
as a single point in the fit.  Stellar effective
temperatures were adopted from Case C of Soderblom et al.\ (1998)
where the single value $T_{\rm eff} = 4350$ K was
given for the spectroscopic binary HD~98800A, and
$T_{\rm eff} = 4250$ and 3700~K were reported for the two stars in the
double-lined spectroscopic binary HD~98800B. Soderblom et al.\ (1998)
reported $A_{\rm V} = 0.44$ mag for HD~98800B, but gave no $A_{\rm V}$
value for HD~98800A. We assumed $A_{\rm V} = 0$ for HD~98800A and used
a standard interstellar extinction law with $A_{\rm V} = 3.1 E_{\rm
(B-V)}$ to redden the model for HD~98800B. The luminosity ratio of the
two components was fixed at 2.7 based on the absolute V magnitudes given by
Soderblom et al.\ (1998) and bolometric corrections from Kenyon \& Hartmann
(1995). The best-fit models gave $L_{\rm star}$ = 0.78 \lsun\ and 0.56 \lsun\
and are plotted as a dotted and dashed line in Fig.\ 2 
for the A and B components, respectively. 

An average dust temperature of 150 K was derived by fitting a
Planck function to the excess continuum emission from 
HD~98800B, omitting points associated with the silicate feature. 
For grains 1-10 $\mu$m in size, 
this temperature corresponds to a 4-12 AU distance 
from a single star of luminosity 0.56 \lsun.
Given the 1 AU orbital separation estimated 
for the components of HD~98800B (Soderblom et al.\ 1998), it 
implies that most of the dust is located in a {\it circumbinary}
configuration around the spectroscopic binary. 
To estimate the radial extent of the dust, we also fit
the spectral energy distribution with a model of a
disk around a single star of luminosity 0.56 \lsun.
The model parametrization and fitting method are described in 
Koerner et al.\ (1998). Five parameters were varied in the fit,  
including inner radius $R_{in}$, radial extent $\Delta R$,
effective particle size $\lambda_0$, and the 
radial power-law index, $\gamma$, of the optical depth,
$\tau(r) = {\tau_0}(r/r_0)^{-\gamma}$. The optical depth scaling, 
$\tau_0$, was derived after varying the area-integrated optical depth,
$\sigma_{total} = {\int^{R_{in}+\Delta R}_{R_{in}}}\  
{\tau_0}(r/r_0)^{-\gamma}\ 2 \pi rdr$, an indicator of the total
cross-sectional area of the grains.
 Parameter ranges considered were 0--9~AU for $R_{in}$, 
1--25~AU for $\Delta R$, $10^{-1}$--$10^3~\mu$m for $\lambda_0$, 
-4.0--4.0 for $\gamma$ and 5--50~AU$^2$ for $\sigma_{total}$.
A disk model with most-likely values of these 
parameters is displayed in Fig.\ 2; these are
$R_{in} = 5.0\pm2.5$ AU, $ \Delta R = 13\pm8$ AU,  
$\lambda_0 = {2}^{+4}_{-1.5}\mu$m, $\gamma = 0\pm2.5$, 
and $\sigma_{total} = 16\pm3$ AU$^2$, where the values quoted are central
within a range of probabilities enclosing the 68\% confidence level. 
The probability distribution is fairly flat within
these ranges and peak values are not always central 
but lie at $R_{in} = 3.0$ AU, $ \Delta R = 22$ AU,  
$\lambda_0 = 1.8\mu$m, $\gamma = 1$, and  $\sigma_{total} = 16$ AU$^2$.

Many of these values are not
narrowly constrained by the flux measurements alone,
largely because the temperature dependence on both
particle size and radial distance from the star makes
it impossible to determine them uniquely.
However, taken over the whole range of parameter space, there
is greater than a 90\% probability that the dust is distributed
in a circumbinary {\it disk}, with $\Delta R/R_{in} > 1$,  
rather than a narrow ring like that around 
HR~4796A ($\Delta R/R_{in} < 0.25$; Koerner et al.\ 1998;
Schneider et al.\ 1999).
We emphasize the caveat that these estimates apply only 
under the assumptions of this particular disk model.
An estimate of the true inner radius, for example, should take into 
account radiation from the two stellar components, and some temperature
broadening may be due to a range of emissivities inherent
in an unknown particle-size distribution.
However, these effects are unlikely to alter our general 
conclusion about the disk vs ring-like nature of the dust.

The total cross section for dust grains around HD98800B,  
$\sigma_{total} = 16\pm3$ AU$^2$, is 2-3 orders of magnitude
smaller than for several other debris disks (e.g., $\beta$ Pic, HR4796A,
and 49 Cet). Thus, from the standpoint of
circumstellar mass, the disk around HD98800B is not as remarkable
as suggested by the infrared excess alone 
(cf. Zuckerman \& Becklin 1993). 
The relatively high fractional luminosity is, instead,
a consequence of dust location 
close to the star where grains intercept a greater 
fraction of the stellar radiation. Assuming a range of plausible grain 
densities, $\rho$ = 1.0-3.0 g cm$^{-3}$, values of $\sigma_{total}$ and 
$\lambda_0$ (grain radius $a$ = $\lambda_0$/1.5; cf.\ Backman et al. 1992)
imply a disk mass in the range of 0.001--0.1 lunar masses.

Models that
incorporate a circumbinary disk surrounding an optically thin
region of warmer dust have served to explain the 
spectral energy distributions of 
younger T-Tauri spectroscopic binaries (Jensen \& Mathieu 1997).
It is likely that HD~98800B is a similar system in 
a later phase of evolution. Modeling of the circumbinary dust emission 
indicates location of the dust in a radial zone
associated with planet building early in the life of our own solar system. 
Consequently, it may well represent the telltale signature of planet 
formation in a hierarchically ordered multiple star system.
If so, we can expect our
picture of the plenitude and diversity of extra-solar planetary 
systems to become increasingly rich as it is revealed by impending
surveys with high-resolution techniques now under development.

\acknowledgments
We gratefully acknowledge support of the NSF's ``Life in Extreme
Environments'' program through grant AST 9714246.
Data presented herein were obtained at the W.M. Keck
Observatory (WMKO), which is operated as a scientific partnership 
among the California Institute of Technology, the University of California 
and the National Aeronautics and Space Administration.  
The Observatory was made possible by the generous financial support of the
W.M. Keck Foundation. 
We wish to thank an anonymous referee for useful comments. A great
debt is due, also, to Robert Goodrich and the WMKO summit staff for their 
many hours of assistance in adapting MIRLIN to the Keck II visitor
instrument port. 

\vfill
\eject


\ \par 
\bigskip
\psfig{figure=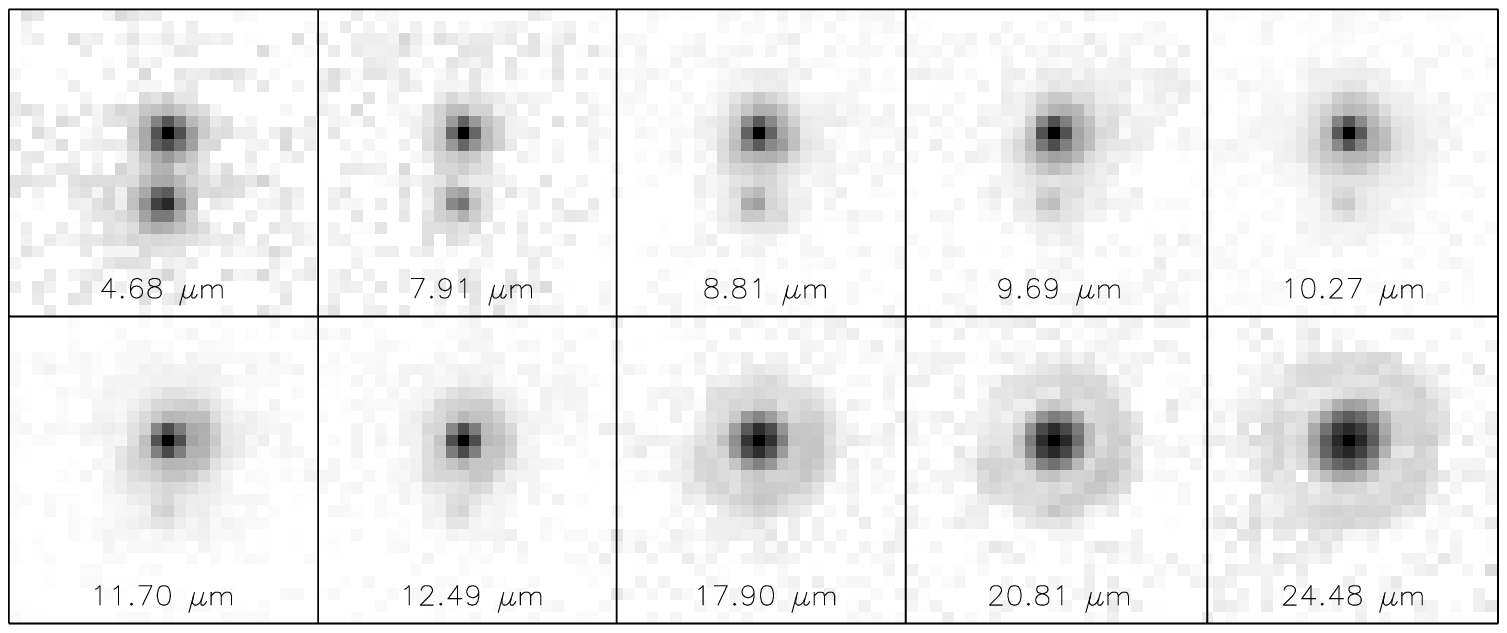,width=6.4truein} 
\medskip

\figcaption{\small Keck/MIRLIN imaging of the thermal infrared emission
from the HD~98800 quadruple system oriented with up axis aligned due North.
The spectroscopic binaries, HD~98800A and HD~98800B, 
are clearly resolved from each other and are identified, respectively,
with northern and southern point sources separated by 0.8$''$ (38 AU). 
Emission from HD~98800A steadily decreases with 
wavelength as $\lambda^{-2}$ and is no longer detected in the 20 $\mu$m 
images. In contrast, radiation from the optical
secondary, HD~98800B, increases dramatically out to 24.5 $\mu$m.}

\bigskip
\vfill
\eject

\psfig{figure=fig2.ps,width=6.4truein,angle=270}

\figcaption{\small Spectral energy distributions for the separate
components of HD~98800. Filled circles in the left plot
are HST (WFPC2 and NICMOS) fluxes for HD~98800A; triangles
represent fluxes listed in Table 1. The dotted line is a
model photosphere from a Kurucz fit to only the HST data. It 
clearly matches the mid-infrared fluxes presented in this work. 
Open squares and diamonds
are plotted in both panels and represent IRAS and JCMT sub-millmeter
fluxes, respectively, for the whole system. Open circles in
the right-hand panel are HST fluxes for HD~98800B; open
triangles are from Table 1. The dashed line
is a model photosphere fit to only the HST fluxes as for HD~98800A.
The dotted and dashed line represents emission from a model disk with
parameters outlined in the text. Together
with the photospheric model, it was fit to measurements at 7.9 $\mu$m,
12.5 $\mu$m, and all longer wavelengths.
Combined photospheric and disk emission is plotted as a solid line.}

\bigskip
\vfill
\eject

\hskip 0.5truein
\psfig{figure=fig3.ps,width=5.0truein,angle=270}

\figcaption{\small Plot of the emission from HD~98800B in the 10 $\mu$m 
silicate band. Filled circles are derived
from flux densities listed in Table 1 by subtraction of continuum
emission interpolated between the points at $\lambda$ = 7.9 and 
12.5 $\mu$m. Vertical error bars are derived from 
uncertainties listed in Table 1. Horizontal error bars refer only
to the filter widths, also given in Table 1. Solid squares
depict the silicate emission feature as observed in the 
interstellar medium toward the Trapezium. Solid triangles and open
circles refer to measurements taken from Hanner, Lynch, \& Russell (1994)
for Comets Levy and Austin, respectively.}

\begin{deluxetable}{lcccccc}
\tablewidth{0pc}
\tablecaption{Component Flux Densities for the  HD~98800 System}
\tablehead{
\colhead{$\lambda_{eff}$} & 
\colhead{$\delta\lambda$} &
\colhead{Calibrator/Flux} & 
\colhead{F$_\nu$(HD~98800)} & \colhead{Component} &
\colhead{F$_\nu$(A)} & \colhead{F$_\nu$(B)} \\
\colhead{($\mu$m)}                &
\colhead{($\mu$m)}                &
\colhead{ \ \ \ \ \ \ \ /(Jy)} &
\colhead{(Jy)}      & \colhead{Ratio (B/A)} &
\colhead{(Jy)} & \colhead{(Jy)} }
\startdata
4.68  & 0.57 & $\beta$ Leo/28.95  & 1.03$\pm$0.090$^1$ 
& 1.21 & 0.47$\pm$0.04 & 
0.72$\pm$0.04 \nl
7.91  & 0.59 & $\beta$ Leo/10.68  & 0.62$\pm$0.03 & 2.46  & 0.18$\pm$0.02  & 
0.44$\pm$0.02 \nl
8.81  & 0.87 & $\beta$ Leo/8.67   & 0.10$\pm$0.04 & 4.88 & 0.17$\pm$0.01 & 
0.83$\pm$0.03 \nl
9.69  & 0.93 & $\beta$ Leo/7.21   & 1.78$\pm$0.07 & 10.20 & 0.15$\pm$0.02 & 
1.62$\pm$0.07 \nl
10.27 & 1.01 & $\beta$ Leo/6.44   & 2.02$\pm$0.08 & 15.38 & 0.12$\pm$0.01  & 
1.90$\pm$0.08 \nl
11.70  & 1.11 & $\beta$ Leo/5.00   & 2.28$\pm$0.09 & 21.28 & 0.10$\pm$0.01 & 
2.18$\pm$0.09  \nl
12.49 & 1.16 & $\beta$ Leo/4.40   & 2.19$\pm$0.09 &  26.32 &  0.08$\pm$0.01 & 
2.02$\pm$0.08  \nl
17.93 & 2.00 & $\alpha$ Hya/42.83 & 4.98$\pm$0.20 & $> 27.78$ &  $< 0.174$ 
& 4.98$\pm$0.27 \nl
20.81 & 1.65 & $\alpha$ Hya/31.74 & 5.53$\pm$0.23 & $> 26.32$ & $< 0.210$  
& 5.53$\pm$0.31 \nl
24.48 & 0.76 & $\alpha$ Hya/30.15 & 8.62$\pm$0.35 & $> 47.62$ &  $< 0.181$  
& 8.62$\pm$0.39  \nl
\tablenotetext{1}{Total flux taken from Zuckerman \& Becklin (1993)}
\enddata
\end{deluxetable}

\end{document}